\documentclass[twocolumn, switch]{article} % Method A for two-column formatting

\usepackage{preprint}

\usepackage{verbatim} %for multiline comment

\usepackage{amsmath, amsthm, amssymb, amsfonts}

\usepackage[numbers,square]{natbib}
\usepackage[utf8]{inputenc}	% allow utf-8 input
\usepackage[T1]{fontenc}	% use 8-bit T1 fonts
\usepackage{xcolor}		% colors for hyperlinks
\usepackage[colorlinks = true,
            linkcolor = purple,
            urlcolor  = blue,
            citecolor = cyan,
            anchorcolor = black]{hyperref}	% Color links to references, figures, etc.
\usepackage{booktabs} 		% professional-quality tables
\usepackage{nicefrac}		% compact symbols for 1/2, etc.
\usepackage{microtype}		% microtypography
\usepackage{lineno}		% Line numbers
\usepackage{float}			% Allows for figures within multicol

\usepackage{lipsum}		%  Filler text

\usepackage{newfloat}
\DeclareFloatingEnvironment[name={Supplementary Figure}]{suppfigure}
\usepackage{sidecap}
\sidecaptionvpos{figure}{c}

\usepackage{titlesec}
\titlespacing\section{0pt}{12pt plus 3pt minus 3pt}{1pt plus 1pt minus 1pt}
\titlespacing\subsection{0pt}{10pt plus 3pt minus 3pt}{1pt plus 1pt minus 1pt}
\titlespacing\subsubsection{0pt}{8pt plus 3pt minus 3pt}{1pt plus 1pt minus 1pt}

\usepackage[utf8]{inputenc} %for french characters
\usepackage[T1]{fontenc}

\newcommand{\eref}[1]{Equation \ref{#1}}
\newcommand{\sref}[1]{Section \ref{#1}}
\newcommand{\fref}[1]{Figure \ref{#1}}

\usepackage{xspace} %to fix \newcommand spacing issue

\usepackage{color, colortbl} %tor table cells highlight

\usepackage{tikz} %for \tikzsymbol

\usepackage{graphicx}[2006/05/08]
\graphicspath{{figures/}}
\DeclareGraphicsExtensions{.pdf,.jpeg,.png,.jpg}

\usepackage{hyperref} %for links to bibliography
\usepackage{amsfonts} %for real set R
\usepackage{soul} %for strikethough text

\usepackage{amssymb} %for \blacklozenge

\usepackage{enumitem}
\usepackage{amsmath} %for multiple lines formula

\title{Fully-Asynchronous Fully-Implicit Variable-Order Variable-Timestep Simulation of Neural Networks}

\usepackage{xwatermark}
\newwatermark[firstpage,color=gray!60,angle=90,scale=0.32, xpos=-4.05in,ypos=0]{\href{https://doi.org/}{\color{gray}{
arXiv:1907.00670
}}}
\newwatermark[firstpage,color=gray!60,angle=90,scale=0.32, xpos=3.9in,ypos=0]{\href{https://doi.org/}{\color{gray}{
arXiv:1907.00670
}}}
\newwatermark[firstpage,color=gray!90,angle=0,scale=0.28, xpos=0in,ypos=-5in]{
correspondence: \texttt{bruno.magalhaes@epfl.ch}
}

\usepackage{authblk}

\author[1]{Bruno R. C. Magalhães}
\author[2]{Michael Hines}
\author[3]{Thomas Sterling}
\author[1]{Felix Schürmann}
\affil[1]{Blue Brain Project, École Polytechnique Fédérale de Lausanne\\Biotech Campus, 1202 Genève, Switzerland}
\affil[2]{Department of Neuroscience, Yale University, New Haven, CT 06510, USA}
\affil[3]{CREST - Center for Research in Extreme Scale Technologies, Indiana University, Bloomington, IN 47404, USA}

\begin{document}

\twocolumn[ % Method A for two-column formatting
  \begin{@twocolumnfalse} % Method A for two-column formatting
  
\maketitle

\begin{abstract}
State-of-the-art simulations of detailed neural models follow the Bulk Synchronous Parallel execution model. Execution is divided in equidistant communication intervals, equivalent to the shortest synaptic delay in the network. Neurons stepping is performed independently, with collective communication guiding synchronization and exchange of synaptic events.  

Commonly to most biological simulations, the interpolation step size is fixed and chosen based on some prior knowledge of the fastest possible dynamics in the system. However, simulations driven by a stiff dynamics  or a wide range of time scales --- such as multiscale simulations of neural networks --- struggle with fixed step interpolation methods, yielding excessive computation of intervals of quasi-constant activity, inaccurate  interpolation of periods of high volatility in solution, and being incapable of handling unknown or distinct time constants. A common alternative is the usage of adaptive stepping methods, however they have been deemed inefficient in parallel executions due to computational load imbalance at the synchronization barriers that characterize the BSP execution model.

We introduce a distributed fully-asynchronous execution model that removes global communication and synchronization, allowing for long variable timestep interpolation of neurons. Asynchronicity is provided by active point-to-point communication notifying neurons' time advancement to synaptic connectivities. Time stepping is driven by scheduled neuron advancements based on synaptic delays across neurons, yielding an \textit{exhaustive yet not speculative} adaptive-step execution.
Execution benchmarks on 64 Cray XE6 compute nodes demonstrate a reduced number of interpolation steps, higher numerical accuracy and lower time to solution, compared to state-of-the-art methods. Efficiency is shown to be activity-dependent, with scaling of the algorithm demonstrated on a simulation of a laboratory experiment.
\end{abstract}
%\keywords{First keyword \and Second keyword \and More} % (optional)
\vspace{0.35cm}

  \end{@twocolumnfalse} % Method A for two-column formatting
] % Method A for two-column formatting

%\begin{multicols}{2} % Method B for two-column formatting (doesn't play well with line numbers), comment out if using method A

%%%%%%%%%%%%%%%  Main text   %%%%%%%%%%%%%%%
% \linenumbers

\begin{figure*}[t]
\centering
\hspace{-0.15cm}\includegraphics[width=1.01\textwidth]{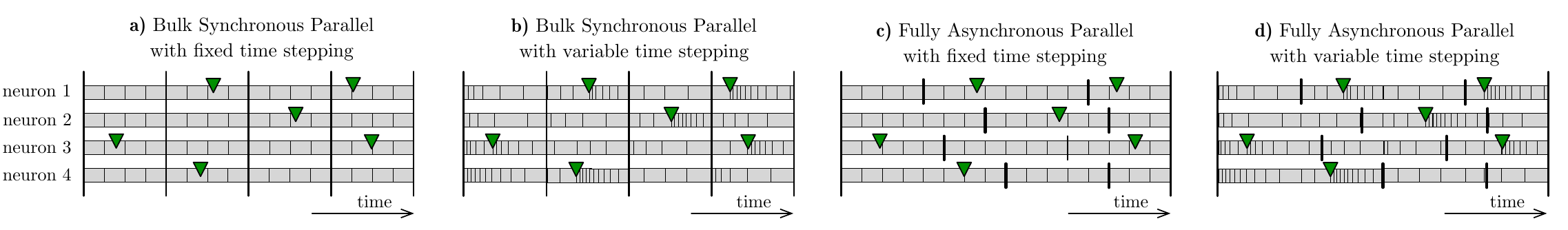}
\caption[]{Scope of the four interpolation methods discussed, illustrated by a left-to-right execution timeline of the simulation of four neurons. A gray cell represents a step of a neuron, with its width representing the timestep duration. Inverted green triangles represent delivery of synaptic events (solution discontinuities). Vertical bars across all neurons represent collective communication. Vertical bars across single neurons represent limit of stepping dictated by pre-synaptic neurons' time instants. 
\textbf{a)} Bulk Synchronous Parallel (BSP) model with fixed timestepping, and collective synchronization; \textbf{b)} BSP model with variable timestep and collective synchronization implemented  in NEURON \cite{lytton2005independent}. Backstepping operations are omitted; \textbf{c)} Fully-asynchronous parallel (FAP) fixed-step method, pioneered by our previous work \cite{magalhaes2019cache}, with individual synchronization of neurons based on their connecting pre-synaptic neurons; \textbf{d)} FAP variable-step method presented in this document.
}
\label{fig_scope}
\end{figure*}

\section{Introduction}

Simulation of the electrical activity of networks of biologically detailed neuron models is a major impact scientific problem, as it allows for the better understanding of the brain. Simulation can be performed on different scales, ranging from molecular-level to point neuron representations, or more detailed neuron models that include neuronal arborization or high number of complex biological mechanisms. Moreover, simulations may focus on different details of neurons activity, from biochemical reactions to conductance-based simulation models, or simpler models such as the integrate-and-fire \cite{brette2007simulation}.  Our problem is the simulation of the electrical activity of large scale morphologically detailed neuron networks. The model of neurons follows from the Hodgkin-Huxley (HH) formalism \cite{hodgkin1952quantitative}, modelling the electrical currents passing though connecting sections of neuron morphologies (spatially discretized as a tree of cylindrical leaky capacitors, henceforth referred to as \textbf{compartments}). Neurons are coupled via electro-chemical transductors, denominated synapses. When the voltage at a neuron soma reaches a specific action potential threshold, a neuron \textbf{spikes} (or \textbf{fires}), leading to a chain of biological reactions that changes the voltage at their synaptically-connected counterparts. 

 Due to the long simulation time required to express biological phenomena such as learning and synaptic plasticity, the acceleration of the simulation of neural networks  is a relevant problem. Similarly to simulations in most scientific fields, existing acceleration efforts follow the Bulk Synchronous Parallel (BSP) execution model, computing several neurons simultaneously via synchronized multithredead and distributed execution. Execution time is divided in communication intervals equivalent to the time duration of the shortest \textbf{synaptic delay} across all neuron pairs in the network, given by the spike propagation delay between a \textbf{pre-} and a \textbf{post-synaptic} neuron. A  synchronous collective communication call performs both synchronization of stepping and synaptic exchange  at the onset of each communication interval. Stepping of neurons is performed independently within the boundaries of each intervals, typically with fixed timestep interpolation, as illustrated in layout a) in \fref{fig_scope}. Such approach is implemented in the NEURON scientific application \cite{hines1997neuron} for the Single Instruction Single Data (SISD) memory layout and in CoreNEURON\cite{kumbhar2016leveraging,CoreneuronArXiv} for the Single Instruction Multiple Data (SIMD, vector-) equivalent. % and arbor \cite{Klijn:836542} simulators.

Further acceleration of individual steps can be achieved on the strong scaling axis with finer-grained parallelism of individual neuron models. State-of-the-art methods provide distributed multicore SIMD acceleration via graph-parallelism of the resolution of the Ordinary Differential Equations (ODEs) \cite{magalhaes2019graph}, and branch-parallelism of neuron topology sections \cite{branchparallelism}. Variable timestep interpolation on the BSP execution model --- pictured in layout b) in \fref{fig_scope} --- is also possible and has been presented by NEURON with SISD computation \cite{lytton2005independent}.  Speculative stepping is allowed by a single step that traverses the synchronization instant, with posterior backstepping for missed events. The step size is limited to the instant of the nearest synaptic or discontinuity event. The main advantage of variable step methods is the dynamic adaptation of the step size to the solution volatility. However, distributed variable-step executions struggle with load imbalance across compute nodes at synchronization barriers and are therefore infeasible for most simulation domains.

Complementary acceleration at the cache level has been pioneered in our previous article \cite{magalhaes2019cache} --- henceforth referred to as \textbf{our previous work} --- where we pioneered the fully-asynchronous parallel (FAP) execution  model applied to fixed timestep interpolations, illustrated in diagram c) in \fref{fig_scope}. The underlying logic is that, because the BSP communication interval refers to the shortest event delay across the system, the removal of collective synchronization barriers and stepping neurons based on their pairwise connectivities allows for stepping intervals longer than the BSP communication timeframe. This promotes better processor prefetching, lower data volume across memory layers, data representations being kept longer in cache, and ultimately, a reduced time to solution.  %The cache-efficient asynchrony requires a fully memory-linear neuron representation, an asynchronous stepping of individual neuron equations based on its time dependencies (supported by an active neuron-to-neuron stepping notification protocol) and a task scheduler that maximises cache locality by tracking neurons progress and advancing the earliest (in time) neuron first. 
%The work introduced for the first time a fully-asynchronous non-speculative fixed-step interpolation protocol, and is illustrated in .

This paper introduces a method for the distributed fully-asynchronous variable-order variable-timestep interpolation of detailed neuron models, that benefits from cache-efficient barrier-free synchronization and performs variable timesteps on the FAP execution model. We show that by following an \textit{earliest neuron steps next} scheduler, we allow for large time interpolation intervals, and maximise the efficiency of the variable step interpolator beyond what was believed to be possible in BSP-based executions. A simple illustration of our methods is presented in layout d) in \fref{fig_scope}.

The contributions of this paper are as follows. We present the mathematical formalism underlying the simulation of our use case and its resolution with variable step interpolation. We perform a study of numerical precision on the electrical activity of a single neuron and demonstrate higher numerical accuracy than its fixed step counterpart found in NEURON.  We analyse and discuss the benefits and bottlenecks of two distinct distributed execution models --- a speculative model with backstepping,  and a scheduled non-speculative model --- and provide insights on their feasibility on distributed simulations of large networks of neurons. We demonstrate a low sensitivity of our model to stiffness of solution (spiking rate), and high susceptibility to discontinuity events (synaptic activity) with a guaranteed speed-up for a discontinuity rate below approx. 1000 Hz. To measure the relevance of our findings on a real use case, we simulate a laboratory experiment based on the spontaneous activity of 219 thousand neurons, demonstrating a mean discontinuity frequency of 94 Hz, and large periods of absence of discontinuities, demonstrating the suitability of our methods to the problem domain.

An implementation of our methods on the core kernel of the NEURON simulator is detailed, and a benchmark is performed on 64 Cray XE6 compute nodes. Distributed asynchrony and multicore executions on a global memory address space is provided by the HPX runtime system \cite{sterling14:_easc}. We simulate five spiking regimes that characterize several dynamics of the mammal brain, on an input of 1024 to 65536 neurons, and compare our methods against five state-of-the-art numerical solvers. Results demonstrate a speed-up of 544-65$\times$ for a quiet spiking regime of 0.25Hz representing a majority of neurons in regular brain activity, down to  7.7-1.8$\times$ to a moderate regime of 6.5Hz, and 2$\times$ to no acceleration for 38Hz, a pattern of unlike occurrence or short duration. An analysis of the overall performance achievable on the previous laboratory experiment demonstrates a speed-up of 224.5-11.9x for an execution with precise delivery of events, increasing to 225.1-17.1x and 228.5-24.6x for two optimized alternatives that group events delivery in the next half and full timestep (a numerical precision similar to the state-of-the-art fixed-step methods). To finalize, we show that over 95\% of neurons fall in three spiking regimes that guarantee the preservation of the speed-up in larger networks, demonstrating good scaling properties of our methods to very large networks of  neurons. 

\section{Methods}

\subsection{Mathematical Model}

\begin{figure}[t!]
\includegraphics[width=0.48\textwidth]{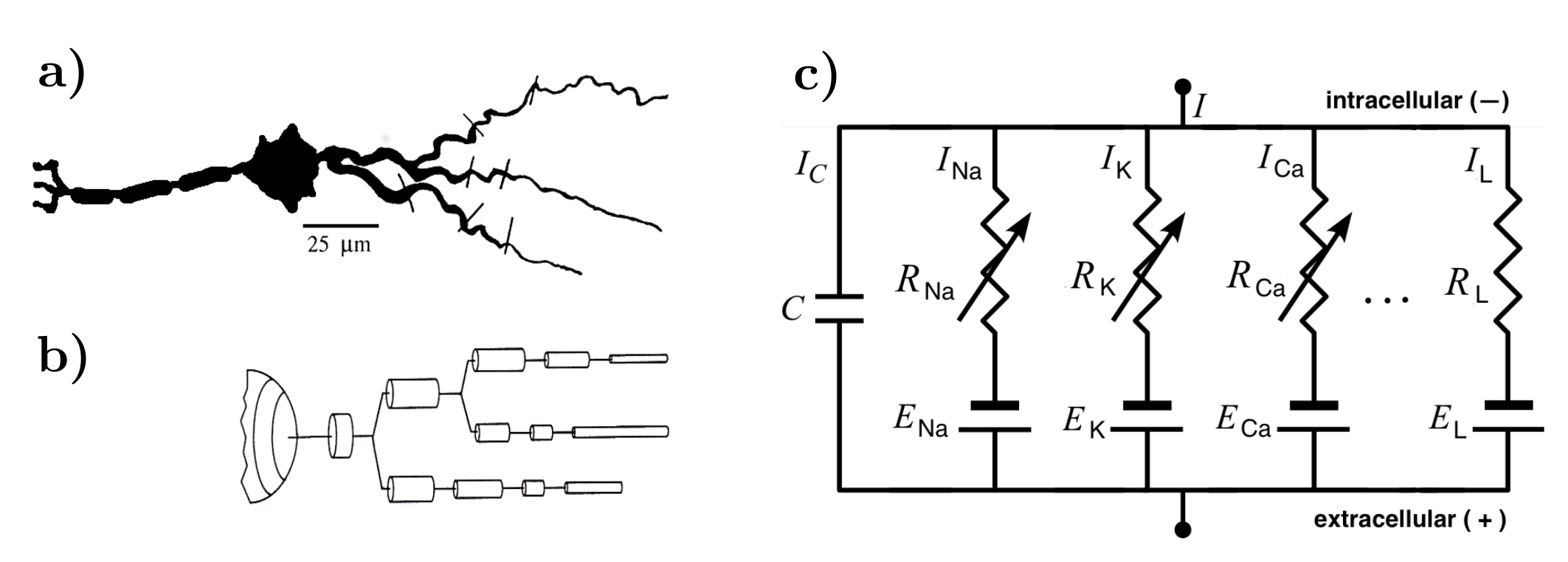}
\caption{\textbf{a)} a sample neuron morphology. \textbf{b)} the spatially discretized model of the compartmental tree of neuron dendrites. \textbf{c)} The RC circuit representing the electrical activity of a compartment represented by the extended Hodgkin-Huxley model, with Sodium ($Na$), Potassium ($K$), Calcium ($Ca$) currents. Remaining currents are included in leak current ($L$).}
\label{fig_neuron_rc}
\end{figure}

The RC circuit that models the electrical current passing through the membrane of a compartment $n$ is modelled as:
\begin{equation}
C \frac{dV_n}{dt} =  - \sum_{i} g_i x_i (V_n-E_i) + I(t)  + \sum_{c : p(c)=n} \frac{V_c - V_n}{r_c } - \frac{V_n - V_{p(n)}}{r_{p(n)}}
\label{eq_current}
\end{equation} 
where $g_i$ and $E_i$ describe the conductance and reversal potential of the ionic channels.  $x_i$ models the opening probability of the transmembrane ion channel currents, typically described by a voltage-gated ODE, and omitted for brevity. Synaptic currents or injected current stimuli, if any, are included in $I(t)$. The branching contributions are provided by Ohm's Law and the neuronal cable theory \cite{ScholarpediaNeuralCableTheory}: the subscript ${p(n)}$ refers to the index of the parent compartment of $n$, and $c$ to an iterator over the indices of its children in the compartmental tree. The variable $r$ defines the axial resistance as a function of the diameter and the cytoplasmic resistivity.  The RC circuit underlying the current passing through a compartment is illustrated in \fref{fig_neuron_rc}, layout c). %The original formulation was introduced by the Hodgkin-Huxley (HH) model, and includes the activity of the ion channel gating states that model the flux of sodium (index Na) and potassium (index K) currents. All remaining currents are included in the leak current (index L), leading to the total current:
%\begin{equation}
%\begin{split}
%\label{eq_HH}
%\sum_{i} g_i x_i (V_n-E_i) & = g_{Na} m^3 h ( V_n-E_{Na}) + g_K n^4 (V_n - E_K)\\&+ g_L (V_n - E_L)
%\end{split}
%\end{equation}
%$m$, $n$ and $h$ are variables describing the opening of the ion channels. Extended HH models include several other ionic states and currents.
%The computational model does not describe capacitance, resistance and conductances as a total value but as a value per unit length instead, and adds the axial resistance of the neurites (between compartments) to the HH-model leading to an equation of the form:
%\begin{equation}
%\begin{split}
%C \Delta x \frac{dV_n}{dt} & =  - \sum_{i} g_i x_i \Delta x (V_n-E_i) + I(t) \Delta x \\ & + \sum_{c : p(c)=n} \frac{V_c - V_n}{r_c \text{ } \Delta x} - \frac{V_n - V_{p(n)}}{r_{p(n)} \text{ } \Delta x}
%\end{split}
%\label{eq_HH_long}
%\end{equation}

%where $r$ defines the axial resistance per unit length as a function of the diameter and the cytoplasmic resistivity. The branching contributions are provided by Ohm's Law and the neuronal cable theory \cite{ScholarpediaNeuralCableTheory}. The subscript ${p(n)}$ refers to the id of the parent compartment of $n$, and $c$ to an iterator over the indices of its children in the compartment tree.  

\begin{figure}[t!]
\centering 
    \includegraphics[width=0.48\textwidth]{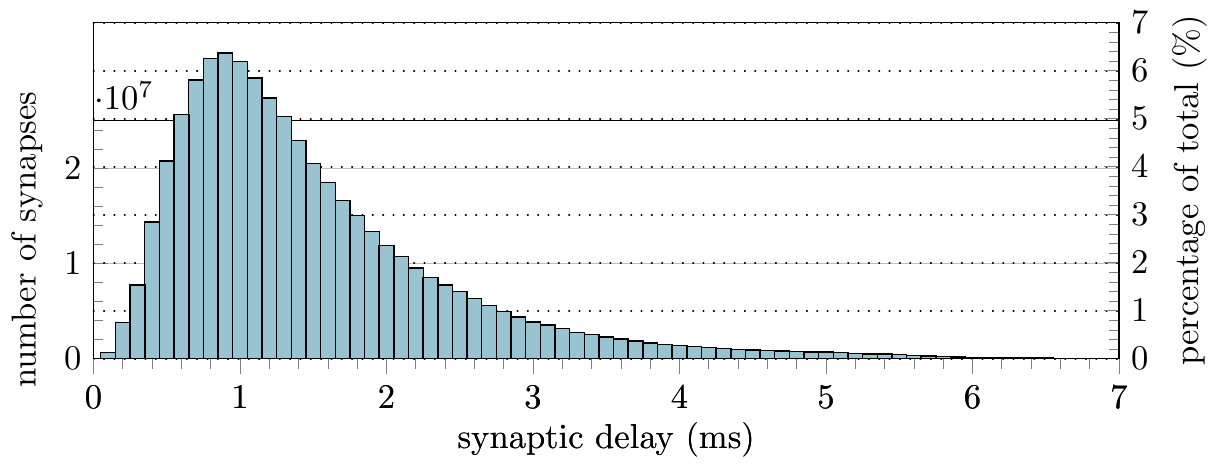}
\caption{Synaptic delays distribution on a network of 219K neurons from the digital reconstruction of the rodent neurocortex from Markram et al. \cite{markram2015reconstruction}. A histogram bin refers to $0.1ms$. The leftmost bar represents the communication step size of current BSP implementations, equivalent to $0.1ms$ and circa $0.13\%$ of the total number of synapses. Intervals larger than $7ms$ (omitted) account for less than $1\%$ of the overall count.}
\label{fig_synaptic_delays_histogram}
\end{figure}

The solution of this system is solved numerically. The complexity of the spatio-temporal model of the neuron activity is reduced by performing a spatial discretization of the neuronal morphology, from biologically inspired to HH-based compartmental representation, and assume the spatial discretization to be small enough, so that the state across compartments' length is constant. Thus, interpolation of solution is performed for consecutive discrete time intervals only.
The resolution follows a fixed step defined as \textit{small enough} to capture the currents with fastest dynamics, set to $0.025$ milliseconds. The fastest synaptic delay across our network model has been measured as $0.1 ms$ or equivalently $4$ computation steps, accounting for $0.13\%$ of the total synapses, as shown in \fref{fig_synaptic_delays_histogram}.

\subsection{Simple and Complex Neuron Models}
\label{sec_simple_complex_models}

A problem specific optimization allows for a substantial speed-up on the resolution of \textbf{simple neuron models} such as the Hodgkin-Huxley, where state variables %$m$, $n$ and $h$
are described by linear ODEs and
depend only on the voltage $V$, and vice-versa. An implicit resolution based on interleaved timestepping of voltage and states, by solving voltages at a given time $t$ and states at time $t+\Delta t/2$, allows for the resolution of the system of ODEs as a system of linear equations.% In practice, one can compute the voltage value at a given time $t$ by utilising the state values precomputed at $t - \Delta t/2$. The converse holds, where state updates utilise the voltage value from the previous half timestep. 

Resolution of \textbf{complex models}, including non-linear and/or correlated state equations cannot be resolved with the aforementioned method, and require a fully-implicit (non-staggered) resolution. A use case of high importance is the model of synaptic plasticity presented by Graupner et al. \cite{graupner2012calcium} --- with cubic ODEs and correlated calcium and synaptic efficacy values --- or Chindemi et al. \cite{GiuseppePlasticity} with higher complexity introduced by ODEs describing synaptic weight and calcium activity models. % present a synaptic model with dependency of state DE $p$ on the calcium current $c(t)$ given by a simple equation and a cubic DE describing the efficacy value $\rho$.  A higher complexity model built on the previous is presented by Chindemi et al. \cite{GiuseppePlasticity}, modelling synaptic weight $w$ and calcium activity $c$ by ODEs.
For such scenarios, numerically reliable resolutions rely on fixed-step iterative implicit methods such as Backward Euler. Alternatively, an implicit variable timestep method with variable order is possible. Its implementation to our use case is detailed in the following section.
 
\begin{figure}[t]
\hspace{-0.2cm}
\includegraphics[width=0.49\textwidth]{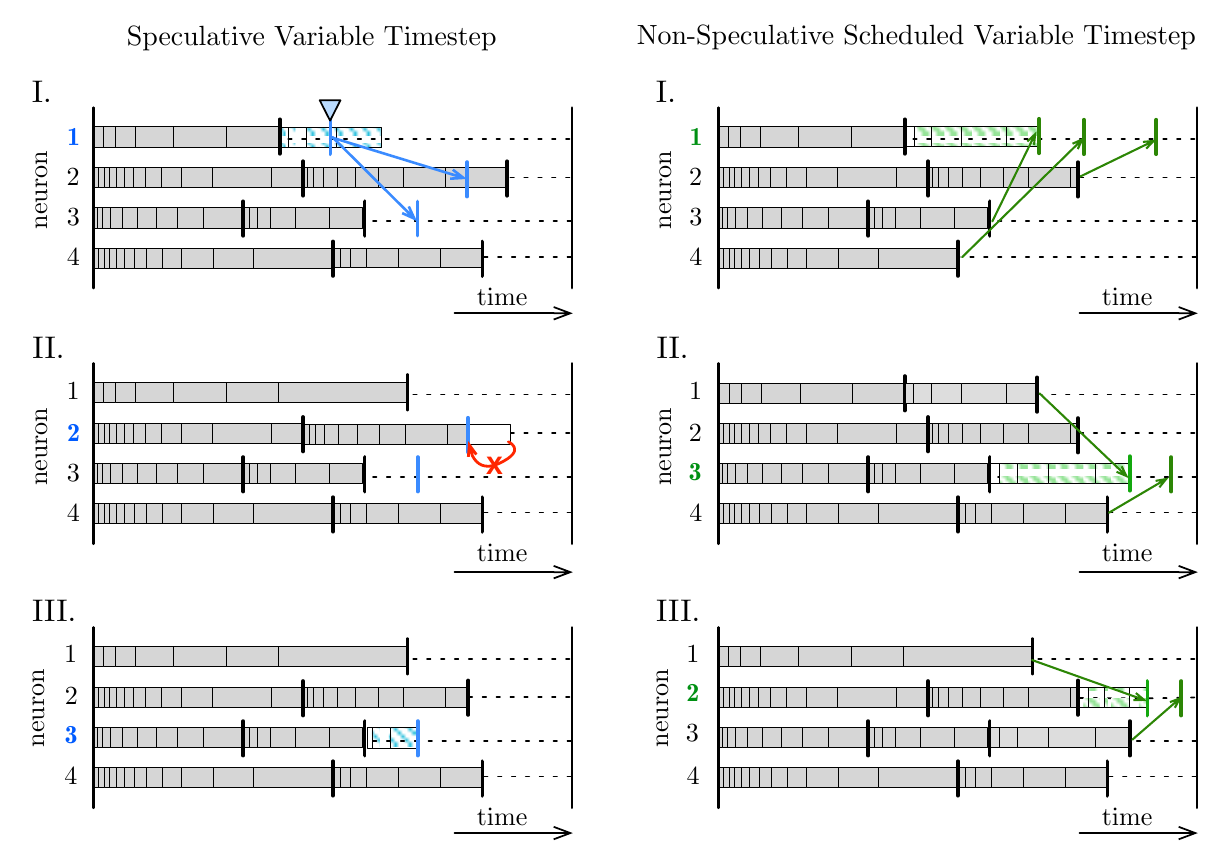}
\caption{Illustrative workflow of two methods for variable timestep (vardt) interpolation. \textbf{Left:} Speculative interpolation advances neurons iteratively, implemented in NEURON on single compute nodes; \textbf{I.} Neuron 1 performs a step (blue area) and spikes during the step interval, with spike timing marked with an inverted triangle. Spike delivery times to post-synaptic neurons 2 and 3 are marked with blue arrow heads; \textbf{II.} The current interpolation time of neuron 2 exceeds the spike delivery time. A back stepping to the previous step follows (red arrow), by interpolating the solution to the instant of the missed synaptic delivery; \textbf{III.} Neuron 3 interpolates until the next (spike) event time. \textbf{Right:} Non-Speculative scheduled time stepping advances the earliest neuron in time, and steps until the next event delivery time or maximum time allowed by pre-synaptic connectivity, to guarantee no backstepping; \textbf{I.} Neuron 1 advances to the earliest time instant allowed (green area), given by the time instant of pre-syn. neurons 2, 3 and 4 and the shortest synaptic delays 2$\rightarrow$1, 3$\rightarrow$1, and 4$\rightarrow$1, respectively  (green arrow heads); \textbf{II.} Neuron 3 is now the earliest neuron in time, and follows analogously based on the synaptic delays of neurons 1 and 4; \textbf{III.} Neuron 2 advances based on neurons 1 and 3. }
\label{fig_backstepping}
\end{figure}

\subsection{Variable Step Implementation}

The CVODE (C Variable-step solver for ODEs, \cite{cohen1996cvode}) is an implementation of the Backward Differentiation Formula (BDF) for the variable-step multistep implicit method solving the Initial Value Problem (IVP, $\dot{y} = f(t,y)$, $y(t_0) = y_0$ where $y \in \mathbb{R}^N$) for ODEs as:
\begin{equation}
\sum_i^{q} \alpha_{n,i} y_{n-i} + \Delta t_n \beta _{n} \text{ } \dot{y} = 0
\end{equation}
where $y = [ ...,V_{k-1}, V_k, V_{k+1},.., x_{i-1}, x_i, x_{i+1}, ...]$ is a vector representing the state variables of a neuron or a set of neurons following the variable notation in \eref{eq_current}, $q$ is the order of the current iteration,  and $\alpha$ and $\beta$ are the $q$-dependent BDF-method coefficients. BDF-1 is the Backward Euler.
%and unfolded as:
%\begin{equation*}
%\small
%\begin{split}
%\text{BDF-1: } & y_n - y_{n-1} = \Delta t_n \text{ } \dot{y};\\
%\text{BDF-2: } & 3y_n - 4y_{n-1} + y_{n-2} = 2 \Delta t_n \text{ } \dot{y};\\
%\text{BDF-3: } & 11y_{n} - 8y_{n-1} + 9y_{n-2} - 2y_{n-3} = 6 \Delta t_n \text{ } \dot{y};\\
%\text{BDF-4: } & 25y_{n} - 48y_{n-1} + 36y_{n-2} - 13y_{n-1} + 3y_{n-4} = 12 \Delta t_n \text{ } \dot{y};\\
%\text{BDF-5: } & 137y_{n} - 300y_{n-1} + 300y_{n-2} - 200y_{n-3} + 75y_{n-4} \\ & - 12y_{n-5} = 60 \Delta t_n \text{ } \dot{y};\\
%\end{split}
%\end{equation*}
%
%Note that the interpolation of order 1 refers to the Backward Euler. 
In brief, CVODE returns the $\Delta t_n$ and $y_n$ that solve BDF-$q$ for an user-provided tolerance \textbf{(atol)}. The computation is performed iteratively, with a suggested step size for each iteration based on the solution gradient and order $q$.
%For a new step $n$, iteration $m$: (1) CVODE suggests $y_{n(m)}$ and $\Delta t_{n}$ based on the user provided jacobian $\dot{y}_n$. The initial guess $y_{n(0)}$ is computed explicitly from history; (2) CVODE provides the Right Hand Side (RHS) computed from the BDF-$q$ formula, and expects an user provided $y_{n(m+1)}$; (3) the implicit resolution relies on Newton iterations, with a stop condition based on an test of $\| y_{n(m)} - y_{n(0)} \| \leq \epsilon$. If error is greater than threshold, a reiteration follows with a smaller $\Delta t_{n(m+1)}$; if error is smaller, proceeds to step $n+1$ with larger $\Delta t_{(n+1)(0)}$.
%The user provides the function that computes the ODE right-hand side for a given value of time $t$ and state vector $y$; and the function that computes the Jacobian $j=\partial f / \partial y$ or an approximation to it.  
The implicit resolution relies on Newton iterations, with a stop condition based on the test $\| y_{n(m)} - y_{n(0)} \| \leq \epsilon$ for iteration $m$ in step $n$: If error is greater than threshold, a reiteration follows with a smaller $\Delta t_{n(m+1)}$; if error is smaller, proceeds to step $n+1$ with larger $\Delta t_{(n+1)(0)}$.  The user provides the function that computes the ODE right-hand side for a given value of time $t$ and state vector $y$; and the function that computes the Jacobian $j=\partial f / \partial y$ or an approximation to it.  We utilise the Jacobian function with the CVODE-defined preconditioning function that solves $P\text{ }x = b$, where $b$ is the input and $P$ approximates $M=I - \gamma \text{ } J$. This is the default Jacobian in NEURON. Two alternative Jacobian implementations were tested and deemed infeasible: (1) a diagonal linear approximation to the Jacobian, given by $J y = \dot{y}$, displayed faster computation yet highly inaccurate results; and (2) a dense matrix approximation $J_{ij} = (f_i (y+h_j, t) -f_i(y,t))/h_j$ for a parameter variation $h_j$, accurate yet infeasible due to the high time to solution and memory requirements for the dense matrix of states.

\begin{figure*}[t]
\centering 
\includegraphics[width=0.217\textwidth]{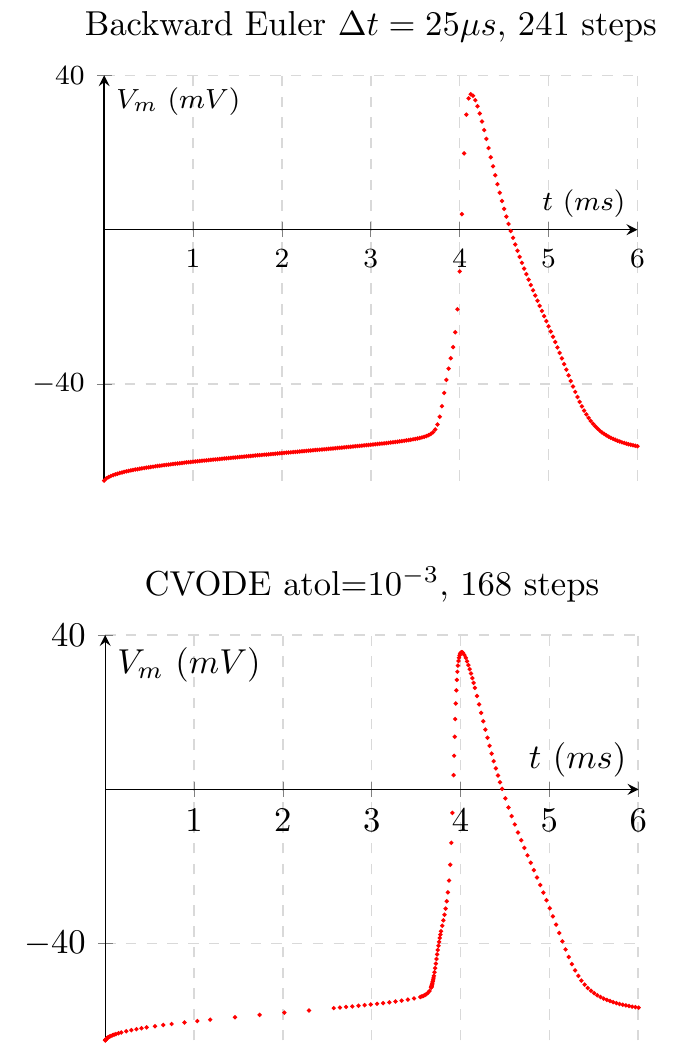}
\hspace{0.01cm}
\includegraphics[width=0.77\textwidth]{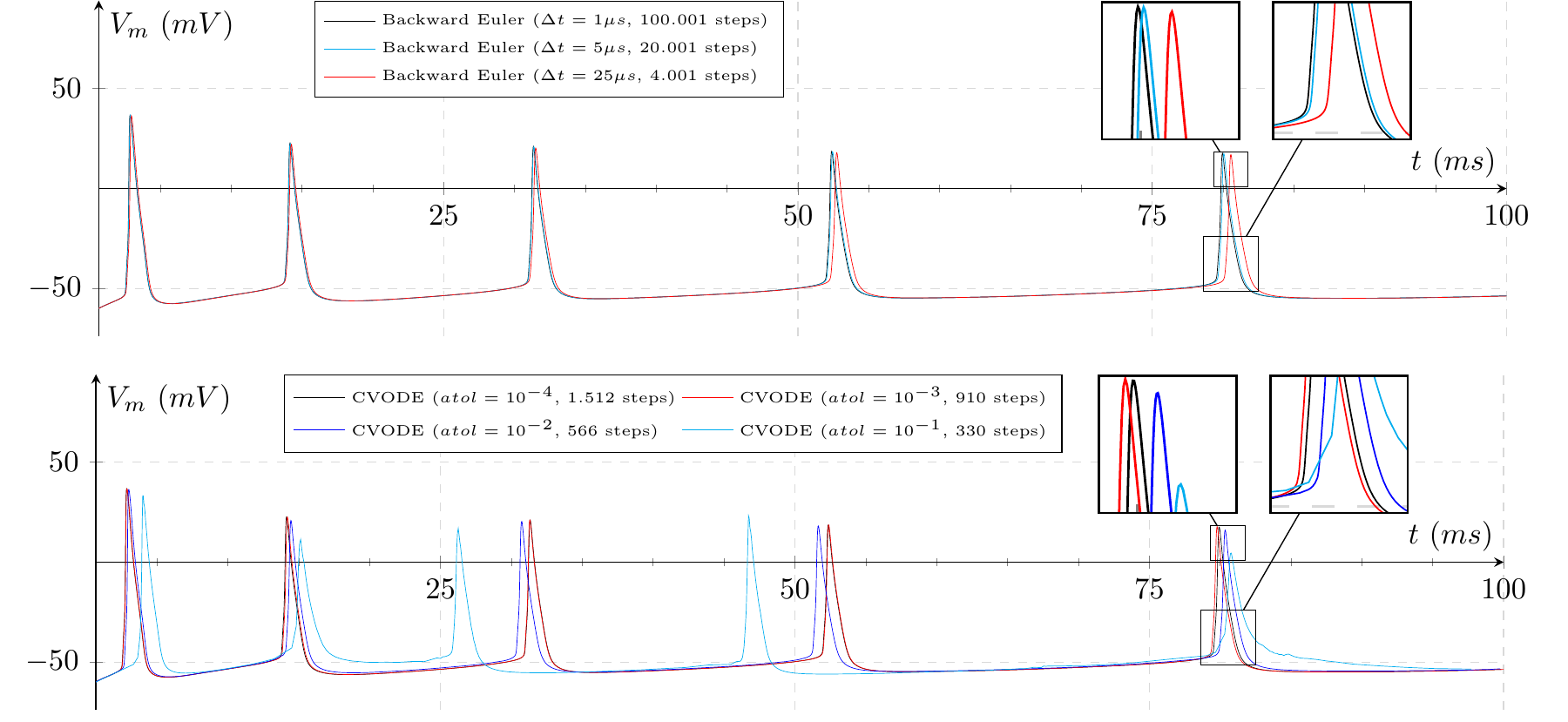}
\caption{Voltage potential at the soma and interpolation steps for a layer 5 pyramidal cell (L5\_TTPC2\_cADpyr232\_1 \cite{NMCportal}) during a $6ms$ action potential voltage trajectory (left) and a $100 ms$ simulation (right)  of a $1.3mA$ continuous current injection, interpolated with Backward Euler (top) and CVODE (bottom) methods, respectively. The reference implementations are the Backward Euler method with $\Delta t = 1 \mu s$ and CVODE with absolute tolerance $10^{-1}$, presented in black and considered indistinguishable. The standard NEURON step size and tolerance values are $\Delta t = 25 \mu s$ and $10^{-3}$ and are presented in red. }
\label{cvode_plots_100ms_currentclamp}
\end{figure*}

\subsection{Asynchronous Timestepping of Neuron Networks}

Control of neurons time advancement on synchronized distributed executions is a solved problem, by enforcing a BSP-like synchronization barrier \cite{hines1997neuron}, as in layouts a) and b) in \fref{fig_scope}. Here we discuss alternatives following an asynchronous execution model. We implemented and analysed the feasibility of two distinct fully-asynchronous barrier-free execution models, displayed in \fref{fig_backstepping} and detailed next.

The initial approach is inspired on the speculative interpolator previously designed for NEURON for a single compute node \cite{lytton2005independent}. Neurons are described by individual interpolators, and advance in time under the best assumption that no \textbf{discontinuity} of solution (synaptic current) will arrive with a delivery time earlier than the neuron current time. Discontinuities lead to a \textbf{reset of the IVP} problem and interpolator state history, and consequently to small steps in the following iterations. When a discontinuity is required to be delivered in a past instant in time, a \textbf{backstepping} operation must precede, in order to reset the recent step and interpolate neuron state back to a time instant of confidence (the time of the discontinuity). Simulations of small neuron networks on single compute nodes are possible and have previously shown a substantial runtime acceleration utilising this model \cite{lytton2005independent}. The state of neurons is local to the compute node, thus backstepping of neuron states and synaptic activity are not computationally heavy and do not require communication. However, in our implementation this approach demonstrated two main drawbacks: (1) large spiking networks lead to a high number of IVP resets, and large amount of time spent on speculative stepping with posterior backstepping. This is mainly due to, in practice, one being unable to tell which neuron should be stepped at a time, such that the risk of backstepping would be minimized and the step length maximized. Moreover, (2) on distributed executions, a main problem arises when synaptic activity (exchanged across different compute nodes) needs to be reverted, requiring further communication. On large networks of neurons, the reversal may lead to an extremely complex cascade chain of reversal notifications and backstepping across neurons on different compute nodes, leading to a high communication and computation workload, deeming this methodology infeasible.

An alternative approach was implemented, based on the non-speculative asynchronous stepping methodology detailed in our previous work\cite{magalhaes2019cache}. Neurons hold a map storing the time instant of their pre-synaptic connectivities. The map is updated by stepping notifications received actively at a certain frequency, throughout the stepping of its pre-synaptic dependencies. The frequency value is set at a neuron pair level, to a minimum value that guarantees no deadlocking. More importantly, this value is at least the BSP communication interval of $0.1ms$, and can reach up to several milliseconds (\fref{fig_synaptic_delays_histogram}), thus minimising overall communication. Neurons step to the maximum time allowed by their synaptic connectivities. This guarantees synapses to be delivered in future time instants, thus removing backstepping and reversion of sent synaptic spikes. The method is improved with an \textit{earliest neuron steps first} scheduler at each compute node that keeps track of neurons advancement, and picks the earliest neuron in time as the next to interpolate. This guarantees the maximisation of the step length and provides a larger variable step interval. Due to the reduced communication and computation, the removal of solution resets and backstepping, and larger stepping intervals, this method will be the default used in the asynchronous variable step simulations of networks of neurons.

\subsection{Implementation details}

Our methods were implemented on the core kernel of the NEURON simulator \cite{CoreNeuronGitHub}. The mathematical specifications of biological mechanisms were provided by the NEURON modelling language (NMODL). SIMD capabilities were implemented on all fixed and variable timestep methods discussed hereafter. Code changes to support SIMD capabilities were added to the MODL-to-C code generator \cite{hines2000expanding}. Communication, synchronisation control objects, memory allocation, threading, distributed execution and parallelism were implemented with HPX \cite{sterling14:_easc}, the runtime system for the Parallex execution model \cite{kaiser2009parallex}. HPX provides a multi-threaded, cooperative thread scheduler, a global address space (GAS), an active-message parcel transport, a group of globally addressable local synchronization object classes, and a dynamic workload scheduler. These features makes HPX very adequate for dynamic adaptive execution of high scalability and high concurrency asynchronous problems on distributed networks of nodes.  The following features were implemented with HPX on a Global Address Space: a memory linearisation of a neurons static and dynamic data structures; an active point-to-point asynchronous communication protocol for stepping and synapse notification; a reduce of communication by aggregating neuron-to-neuron into into neuron-to-compute node communications; and a neuron scheduler tracking neurons' time advancement and dynamically allocating computing resources to the earliest neuron in time. Implementation details have been covered in our previous manuscript \cite{magalhaes2019cache}, and will be omitted for brevity.

\section{Results}

\subsection{Numerical Accuracy}

Backward Euler is an A-stable and L-stable method of order 1. The numerical accuracy of CVODE depends on the order of the Backward Differential Formula iteration and the user-provided tolerance. The underlying BDF  is A-stable at order 1 or 2. At orders 3 to 5, it is not A-stable but \textit{stiffly-stable}: the region of instability grows as the order increases from 3 to 5, and stability depends on the step size \cite{cohen1996cvode}.

We compare the numerical accuracy of both models by measuring the time difference of the main unit of interest in the activity of spiking neuron networks --- the spiking time instants. \fref{cvode_plots_100ms_currentclamp} presents the voltage trajectory and number of steps of a $1.3mA$ current clamp experiment for a single ($6ms$) and several ($100 ms$) spikes of a layer 5 pyramidal cell. The default settings utilised in NEURON are plotted in red, with $\Delta t = 25 \mu s$ and $atol=10^{-3}$ for Euler and CVODE methods, respectively. The reference solutions are plotted in black, with $\Delta t = 1 \mu s$ and $atol=10^{-4}$, and their numerical difference is considered negligible.

Results on the single spike voltage trajectory ($6ms$) display a reduced step count and better adaptation to trajectory change when comparing CVODE to Euler method. The rationale behind the better performance of adaptive stepping is that it is gradient sensitive and thus it follows the natural behaviour of a neuron: a cell spike requires small timesteps for higher precision, followed by interspike intervals or resting periods with little synaptic input therefore allowing large step sizes. CVODE displays less steps during long periods of low gradient (e.g. $1-2.5ms$), and greater number of steps for steep trajectories (the uprising trajectory of the spike) and sudden changes in gradient (the trajectory proximal to the peak voltage).

\begin{figure}[t]
\hspace{-0.2cm}
 \includegraphics[width=0.585\textwidth]{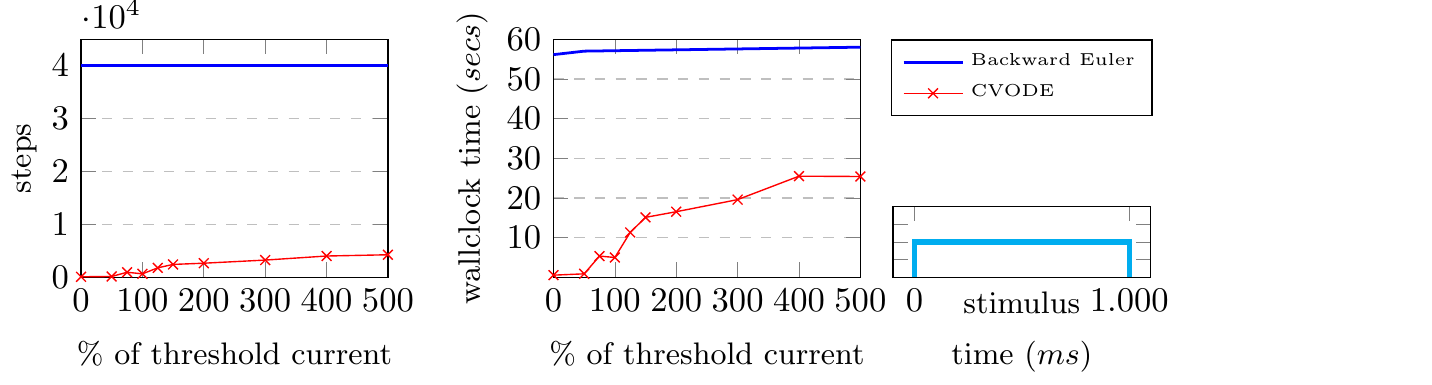}
\caption{Interpolation steps and runtime for the $1000ms$ simulation of the injection of a continuous current as a percentage of the threshold current ($0.206 mA$) on a layer 5 pyramidal cell (L5\_TTPC2\_cADpyr232\_1 \cite{NMCportal}). Results presented for the Backward Euler with $\Delta t = 25 \mu s$ and CVODE with atol=$10^{-3}$. Hardware specs.: Intel core i5 at 1.6 GHz.}
\label{cvode_dependency_variation_solution}
\end{figure}

 The $100ms$ simulation studies the impact and numerical accuracy of both interpolators to longer executions. Results display a phase shift in solution (measured as the time difference between peak voltage values of the reference and benchmark curves) that increases with the increase of the step size on the Euler methods. In practice, the timestep determines the fastest reaction time of the system. Thus, a large timestep will inevitably cause the system dynamics to be \textit{slow}. The analysis of the performance of the CVODE tolerance values show that tolerance of $10^{-2}$ approximates the resolution of the default Euler step size of $25 \mu s$, with a significant reduction of $7\times$ in step count. At longer runs, the variable step demonstrated to be more precise, due to no accumulation of phase shift, with the maximum trajectory shift measured at approximately $1.1ms$. On the other hand, a tolerance value of $10^{-3}$ approximates closely the optimal solution with $40\%$ less steps, and with a margin of error similar to its $5 \mu s$ Euler counterpart for the period of $100ms$, while yielding $22\times$ less interpolations. %We assume these CVODE-Euler performance improvements to be best case scenarios, as the accuracy of the CVODE method is activity-dependent and in this scenario there are no discontinuities in the system leading to resets of IVP.

On a longer execution, the measured phase shifting of the trajectory for a period of $1000ms$ of simulation (omitted for brevity) was of approximately $2 ms$ for $\Delta t = 5 \mu s$, and $6.5 ms$ for $\Delta t = 25 \mu s$, considered a large value in the timescale of neurons activity. The $100ms$ execution displayed demonstrated a phase shifting of approximately a tenth of the whole second simulation. 

\subsection{Performance Dependency on Solution Stiffness}

We measured the response of both stepping methods with the  NEURON default parameters to different levels of changes in spiking frequency. Changes in trajectory were enforced by injecting a continuous current of a given amplitude on a neuron during $1000ms$. Current intensity is measured as a percentage of the \textbf{threshold current}, the minimum continuous current value that needs to be injected to force the neuron to spike. Performance was measured in terms of steps count and time to solution on an Intel i5 at 1.6 GHz. Results are presented in \fref{cvode_dependency_variation_solution}, and demonstrate that high dynamics of the solution degrade the CVODE performance. This is justified by CVODE requiring smaller steps on high trajectory variations in order to respect the absolute tolerance value. For the range of tested scenarios, CVODE runs on less steps and shorter time to solution when compared to Backward Euler. For a neuron without any spikes, or equivalently for a current injection below the spiking threshold, it was measured a reduction of: (1) $434\times$ in step count and $98\times$ in runtime for injected currents below $50\%$; (2) $62\times$ step count and $11.6\times$ runtime for $100\%$ of the threshold current; and (3) $9.4\times$ step count and $2.5\times$ runtime for $500\%$ of the threshold current, a worst case scenario of little probability of occurrence.

\subsection{Performance Dependency on Number of Discontinuities}
\label{sec-discontinuities}

\begin{figure}[t]
\hspace{-0.2cm}
 \includegraphics[width=0.6\textwidth]{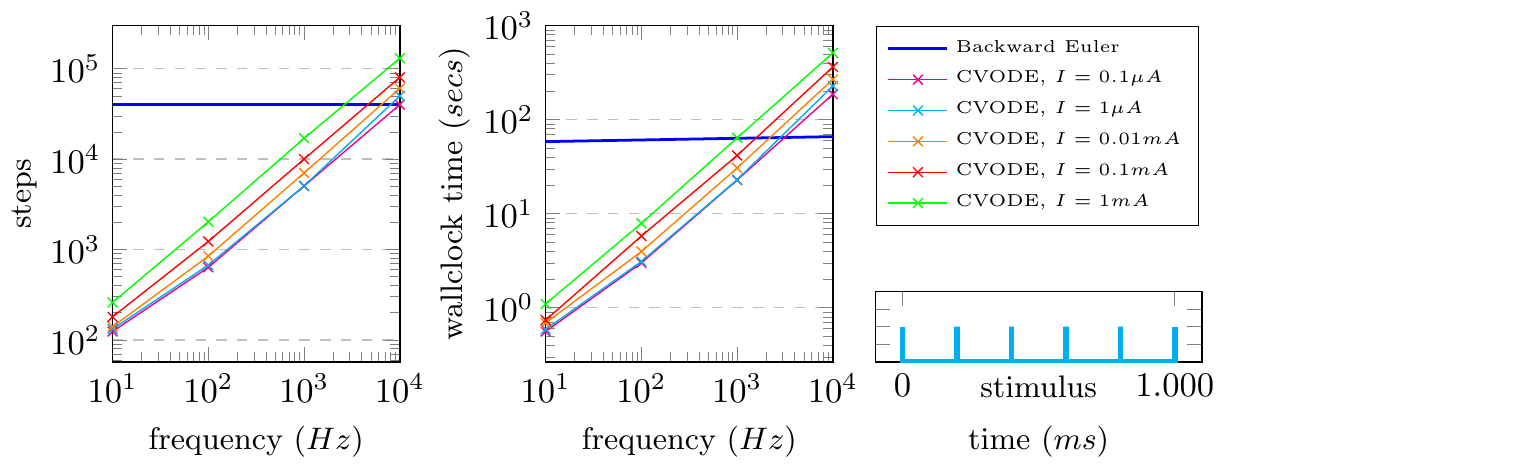}
\caption{Interpolation steps and runtime for the $1000ms$ simulation of the injection of short $1 \mu s$ current pulses of different amplitudes $I$ at different frequencies, on a layer 5 pyramidal cell (L5\_TTPC2\_cADpyr232\_1 \cite{NMCportal}). Results presented for the Backward Euler with $\Delta t = 25 \mu s$ and CVODE with atol=$10^{-3}$. Hardware specs.: Intel core i5 at 1.6 GHz.}
\label{cvode_dependency_events_arrival}
\end{figure}

\begin{figure}
\raggedright
\small a) \vspace{0.1cm}\\
\includegraphics[scale=0.88]{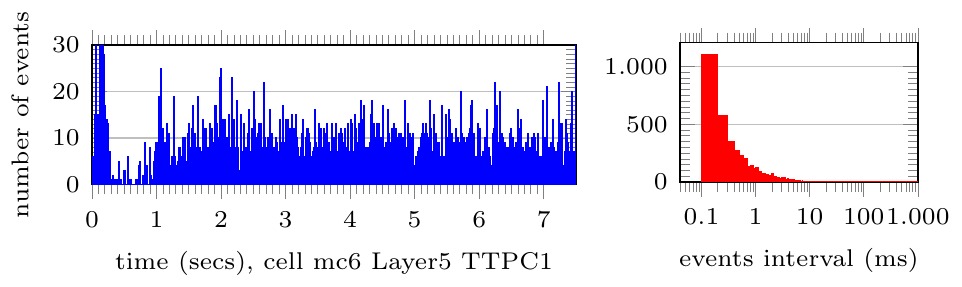} \\
\includegraphics[scale=0.88]{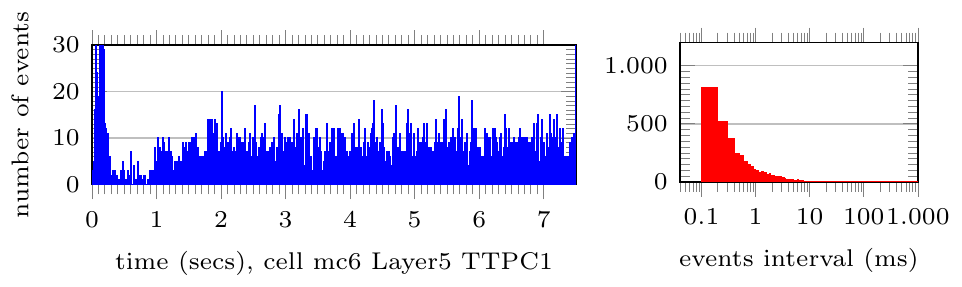} \\
\small b) \vspace{0.1cm}\\
\includegraphics[scale=0.95]{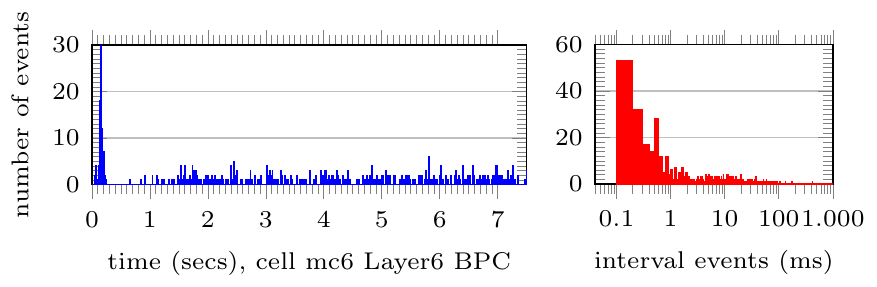} \\
\includegraphics[scale=0.95]{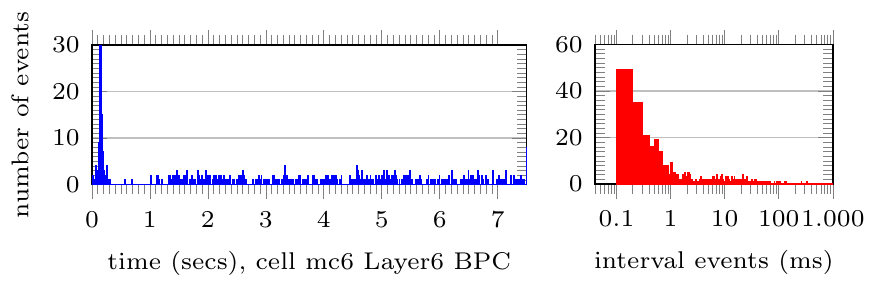} \\
\small c) \vspace{0.1cm}\\
\includegraphics[scale=0.95]{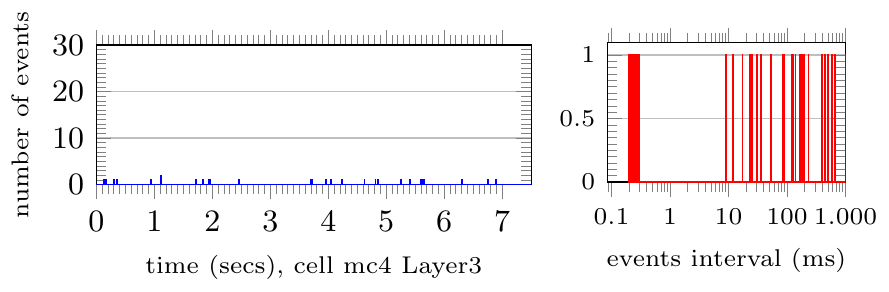} \\
\includegraphics[scale=0.95]{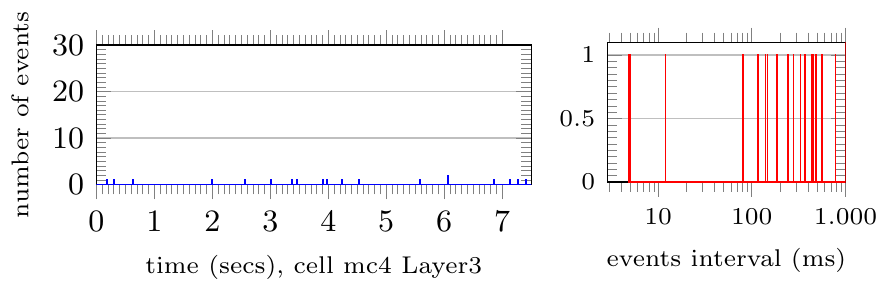} \\
\caption{Number of discontinuity events (incoming spike currents, bin size 0.25ms, left) and distribution of events time difference (bin size 0.1ms, right) throughout 7.5 secs of simulation, for two sample neurons collected from the \textbf{a)} top 1\% neurons, receiving between 3040 and 6146 events; \textbf{b)} median 1\%, from 541 to 558 events; and \textbf{c)} bottom 1\% of neurons receiving less than 100 events. Total events count: circa 155 Million; mean per neuron: 707.
}
\label{fig_events_O2_circuit}
\end{figure}

We measured the effect of discontinuities on both methods by injecting several current pulses at a fixed frequency on a neuron soma, in the same Intel i5 compute architecture, mimicking synaptic events. The experiment results are displayed in \fref{cvode_dependency_events_arrival} and suggest that the CVODE performance depends on the trajectory change from each discontinuity, i.e. the amplitude of the current injected. This is due to the increase of the voltage trajectory being dependent on the amount of current injected. The larger the voltage increase, the larger the change in trajectory gradient, thus the more interpolation steps are required. Furthermore, and as expected, results demonstrate that the number of discontinuities plays a major role on the CVODE performance. CVODE is shown to deliver a reduction of steps in the order of $153-322\times$ for a frequency of 10 discontinuities per second  for the current values of $1 mA$ to $0.1 \mu A$. The step count equilibrium between Euler and CVODE method lies in the interval of $10^{3.2}-10^{4.0} Hz$ for the same currents interval.

The runtime demonstrates a similar dependency on the injected current, yielding a speed-up of $51\times$ for $10Hz$ decreasing linearly up to the speed-up equilibrium value at $1000Hz$ for the strongest current. For the lightest current injected, a speed-up of $100\times$ is visible for the $10Hz$ discontinuity rate, decreasing to an Euler matching value at circa $1600$ events per second ($10^{3.2} Hz$). Although this exercise does not represent the stochastic pattern of spikes arrival on real neurons, typically described by a Poisson distribution with a long tail, it provides an estimation of the CVODE speed-up allowance. With that in mind,  the following section measures the discontinuity rates on a simulation of a laboratory experiment.

\subsection{Simulation of a Laboratory Experiment}
\label{sec-in-vivo-exp}

We tested the suitability of variable step methods to our problem by measuring the spiking activity of a simulation of 7.5 secs of electrical activity mimicking a laboratory experiment. The experimental set-up performs a fixed step simulation of the spontaneous activity of 219.247 neurons during tonic depolarization. The network exhibits spontaneous slow oscillatory population bursts, initiated in layer 5 (L5), spreading down to L6, and then up to L4 and L2/3 with secondary bursts spreading back to L6. For further details, refer to section \textit{Simulating Spontaneous Activity} in \cite{markram2015reconstruction}.

A representative distribution of discontinuity events for three groups of neurons --- organized by highest 1\%, median 1\%, and lowest 1\% number of discontinuities --- is displayed in \fref{fig_events_O2_circuit}. The simulation incurred a total of circa 155 million events, with the following distribution: (a) top 1\% of neurons, between 3040 and 6146 events in 7.5 secs, or 405-820 Hz; (b) median 1\%, from 541 to 558 events (72-74.4 Hz); and (c) bottom 1\%: less than 100 events ($\leq$10 Hz).  The average number of events was of 707 events for the 7.5 secs of simulation, or equivalently, 94 Hz, significantly below the $1000 Hz$ threshold discussed in the previous section. Moreover, the results on the distributions of time interval between discontinuities, plotted in red on the right, display large periods of \textit{silence} between events arrival in the median and bottom use cases, but not on the top, suggesting the suitability of adaptive stepping to most (but not all) neurons in the population.

\begin{figure*}[t]
\centering 
\includegraphics[width=1.01\textwidth]{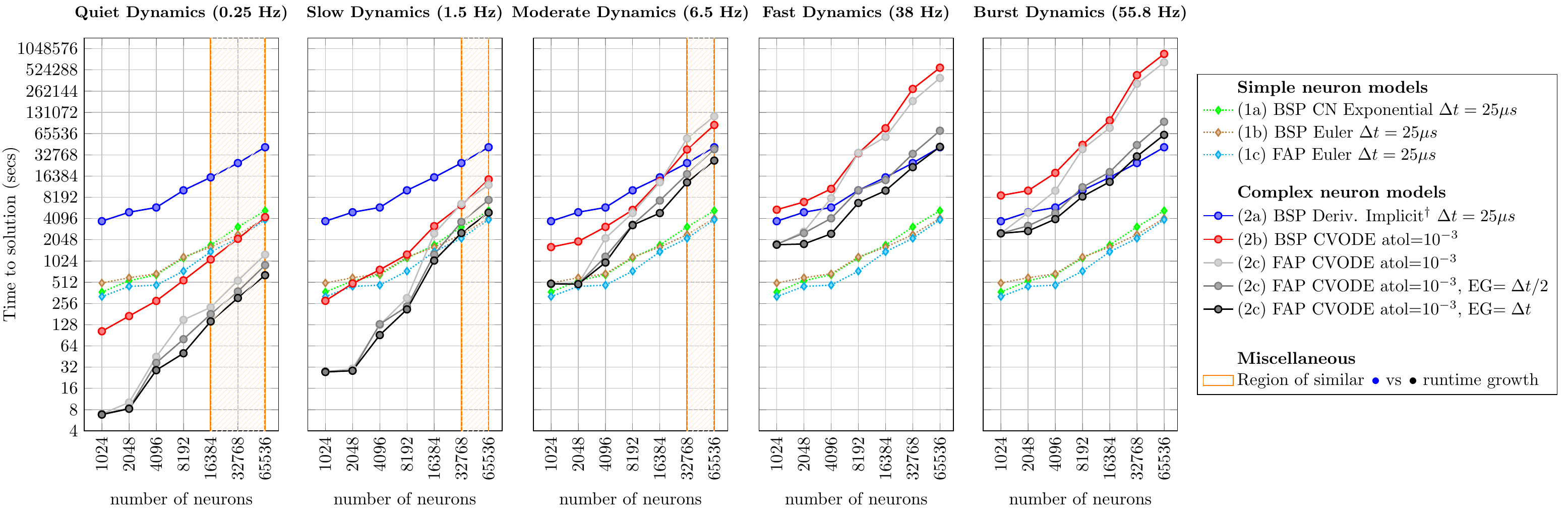}
\caption{Runtime for the simulation of one second of biological activity of neuron networks described by five spiking rate dynamics. Results measured for increasing input sizes on a network of 64 Cray XE6 nodes, with an AMD Opteron 6380 with 16 cores at 2.5 GHz. \textbf{Key:} BSP: Bulk Synchronous Parallel; FAP: Fully-Asynchronous Parallel; atol: absolute tolerance; EG: event grouping interval;  $^\dagger$: able to solve non-linear ODEs implicitly, and unable to solve correlated mechanism states implicitly.}
\label{fig_benchmark_xe6}
\end{figure*}

\section{Benchmark}
 
%In our previous work we demonstrated scale and cache efficiency for the fully-asynchronous methods on the resolution of simple neuron models via fixed step interpolation. Results presented: (1) reduction of inter-layer cache  bandwidth due to higher number of steps in cache and linear data structures; (2) larger stepping intervals when performing scheduled stepping of earliest neuron in time; (3) a speed-up of up to 65\% in time to solution across four distinct compute architectures --- 18-core Intel Xeon 6180, 64-core Intel Knights Landing, 16-core Intel E5 and 16-core AMD Opteron; and (d) a speed-up of up to 40\% on 32 AMD Opteron (Cray XE6) compute nodes. For brevity, the following study will focus on the netnetworks of executions via the fully-asynchronous variable timestep interpolation.
We simulate one second of the electrical activity of a digitally reconstructed neural network extracted from the model of Markram et al. \cite{markram2015reconstruction}.  Neuron models include 23 distinct biological mechanisms types modelled by 44 ODEs, and highly heterogeneous neuron morphologies.  Each neuron requires a storage of 5-15 MB for its state, times $q$ for a $q-$order interpolation. Our CVODE solver is defined to utilise the default maximum BDF order value of 5.

On the set-up of the simulation test bench,  it is relevant to mention that neuronal activity is highly dependent on the mammal specie, brain region and momentary activity, among other factors. Furthermore,  the network behaviour simulated must approximate a real use case, as spiking activity affect heavily the performance of variable step methods, as detailed in \sref{sec-discontinuities}.  An analysis of a single simulation combining several brain dynamics would be of little interest in the context of efficiency analysis, due to all free variables that affect performance. Therefore, our test bench simulates and studies the efficiency of five different brain dynamics described in the literature:
\begin{enumerate}[leftmargin=13pt]
\item  a model of \emph{quiet dynamics} with a mean spiking rate of 0.25 Hz per neuron, representing neurons almost at rest and/or with little activity. This model provides an upper bound of the runtime of circa 90\% of neurons in the brain during regular activity --- see Table 1 in Shoham et al. \cite{shoham2006silent} for evidence of silence and highly sparse activity among neurons; supported by Kerr et at.  \cite{kerr2005imaging} for estimates of rat's neocortex spiking rate at circa 0.1 Hz due to sparsity of activity; and the estimation of 0.16 Hz by Lennie et al. \cite{lennie2003cost} based on brain energy levels;
\item a model of \emph{slow dynamics} at 1.5Hz, representing the lower bound of active neurons, described next;
\item a model of \emph{moderate dynamics} with a spiking rate of 6.5 Hz, an approximation of the irregular regime of slow oscillations displayed by the Brunel network \cite{brunel2000dynamics}; also an upper limit to the 0.005-5Hz spiking rate of the rate frontal cortex \cite{watson2016network}; and a lose approximation of the to visual cortex of the cats in the 3-4 Hz interval\cite{baddeley1997responses}.
\item a model of \emph{fast dynamics} of 38 Hz, in the range of 30-40 Hz characterizing cortical and thalamic neuronal activity during periods of high vigilance \cite{steriade1998dynamic}; and the regular regime of the theoretical model of inhibition-dominated model of the Brunel Network \cite{brunel2000dynamics}; and 
\item a model of \emph{burst dynamics} at 55.8 Hz, typically a by-product of depolarizing current injections, similar to the first instants of simulation in \fref{fig_events_O2_circuit}; and representative of the fast spiking regime of the  inhibition-dominated irregular dynamics of the Brunel Network \cite{brunel2000dynamics}.
\end{enumerate}

The following benchmarks measure the scaling of our methods by simulating quasi-homogeneous activity across neurons on densely connected networks. This set-up is favourable to fixed step and BSP-based methods, therefore the results presented next are a lower bound of possible acceleration. Neurons activity is triggered by a constant current injection in all neurons throughout the whole duration of the simulation, strong enough to approximate the spiking rate of the network to the regimes described. For complete coverage of the topic, we include the following state-of-the-art solvers for simple neuron models (labelled 1a to 1c, and restrained to linear ODEs with uncorrelated states) and complex models (2a-2c), both detailed previously in \sref{sec_simple_complex_models}:
\begin{description}[leftmargin=20pt,labelwidth=0.5cm]
\item [1a)] the \textit{cnexp} fixed step solver in NEURON, with added SIMD, providing an intercalated resolution of current and states as linear equations, with an analytical resolution of the first order ODE describing state variables; 
\item [1b)]  the \textit{Euler} solver in NEURON, with added SIMD, resolving the current-states dependency with an explicit Euler method with staggered timestepping, and as a linear equations; a model computationally less expensive than the previous due to no exponential and division operator;
\item [1c)] the same \textit{Euler} method on a fully-asynchronous parallel execution model, presented in our previous work, and illustrated in diagram c in \fref{fig_scope};
\item [2a)] the BSP fixed step \textit{derivimplicit} solver available in NEURON, with added SIMD, with interleaved-timestep resolution of current as a linear equation, and implicit resolution of individual mechanism state ODEs;
\item [2b)] the BSP variable step method in NEURON with added SIMD and a collective communication barrier, show in diagram b) in \fref{fig_scope}; and
\item [2c)] the SIMD-enabled fully-asynchronous protocol with variable timestepping introduced in this paper (diagram d in \fref{fig_scope}).
\end{description}

The following biological constraints were taken into account to guarantee that variable-step runtimes represent a biologically plausible use case: we verified that there were no continuous periods of silence in the network, and no collective synchrony of spiking or voltage trajectory across neurons, that would promote additional efficiency in variable timestep methods. Our input data is retrieved from neurons in layers 4 and 5 of the brain, typically represented by the longest dendritic trees, therefore representing a worst case scenario in terms of number of pre-synaptic (dependency) neurons. Finally, the number of synapses as AMPA and GABA receptors was measured, and are characterized by a mean of 2289.7 and 4418.8, and a std. dev. of 1284.9 and 3200.4 per neuron, below the counts described in the literature, but inline with the reference digital reconstruction of the rodent brain detailed in \sref{sec-in-vivo-exp}.

Execution times were collected on 64 Cray XE6 compute nodes, powered by an AMD Opteron 6380 with 16 cores at 2.5 GHz, 64 GB of RAM and 256-bit floating point units. Efficient point-to-point communication and remote direct access memory is provided with specialized Infiniband network hardware, interfaced via the photon API library \cite{kissel2016photon}. 

To study the dependency of the algorithm on the input size and synaptic connectivity, we tested our methods in neural networks ranging from 1024 to 65536 neurons, a scale that approximates two columns in the rodent neocortex, and the maximum allowed due to memory requirements of the BDF solver of order 5 we used. Neurons were equally distributed in a round robin fashion across compute nodes. Due to the long simulation time required, runtimes for moderate, fast and burst dynamics were extrapolated from an execution of $250$, $100$ and $100 ms$, respectively.

The benchmark results are presented in \fref{fig_benchmark_xe6}. The FAP variable step method (2c$\hspace{0.3mm}{\color{lightgray}\bullet}$) is presented alongside two variants --- labelled 2c$\hspace{0.3mm}{\color{gray}\bullet}$ and 2c$\hspace{0.3mm}\bullet$ --- that group and deliver instantly the discontinuity events within an interval equivalent to the timestep $\Delta t/2$ of the interleaved fixed timestep method, and the $\Delta t$ of the regular fixed step methods, respectively. This approach yields a level of reduced precision in the delivery of events --- similar to fixed step methods --- however maintaining the same high variable-order variable-step accuracy during periods of activity without discontinuities, with the main advantage of reducing significantly the number in IVP resets. CVODE-based executions are displayed for an absolute tolerance (atol) of $10^{-3}$, a value equal to the default value in the NEURON simulator. Executions with an absolute tolerance of $10^{-2}$ yielded a reduction in runtime of 5\%-8\%, and were omitted for brevity.  The performance analysis for fixed-step simple model solvers  (1a-1c) was covered in depth in our previous work \cite{magalhaes2019cache}, and is omitted for brevity. 

\subsection{Fixed- vs Variable-Timestep Interpolators}
\label{results_fixed_var}

Fixed step methods do not yield significantly-different execution times across different spiking regimes. This is due to the homogeneous computation of neuron  state updates throughout time, and the light computation attached to synaptic events and collective communication not yielding a substantial increase of runtime. The difference in execution times measured across the five regimes was of about $2\%$, which we consider negligible. On the other hand, as expected, variable step executions are penalized on regimes with high discontinuity rates. It is noticeable that the runtimes of fixed- and variable-step solvers approximate as we increase the spiking rate, i.e. the increase of runtimes with the input size is steeper for variable timestep (2b$\hspace{0.3mm}{\color{red!80}\bullet}$ and  2c$\hspace{0.3mm}{\color{black}\bullet}{\color{gray}\bullet}{\color{lightgray}\bullet}$) compared to fixed timestep methods (2a$\hspace{0.3mm}{\color{blue!80}\bullet}$). This is due to discontinuities in variable-step being delivered throughout a continuous time line, compared to the discrete delivery instants of the fixed-step methods --- therefore increase the number of interpolation steps; and the iterative model of the variable timestep reinitializing the state computation with small step sizes on each IVP reset, compared to the constant-sized step of fixed step methods.  A remarkable performance is visible on the quiet dynamics use case, where our fully-implicit ODE solver of complex models (with Newton iterations), still runs faster than the simple solver resolving only a system of linear equations. The underlying rationale is that --- despite the inherent computation cost of Newton iterations in the variable step methods --- the low level of discontinuities allow for very long steps, that surpass the simulation throughput of simple solvers running on fixed step methods. 
The measured speed-up of our reference method (2c$\hspace{0.3mm}{\bullet}$) compared to the reference fixed step method (2a$\hspace{0.3mm}{\color{blue!80}\bullet}$) was of 544-65$\times$ across input sizes for the quiet dynamics, down to  7.7-1.8$\times$ to the moderate dynamics.  The fast dynamics presented a speed-up of twofold for the dataset of 1024 neurons, and a similar runtime for the 66K neurons. The burst dynamics, although of very unlikely probability of occurrence, demonstrated an acceleration of 1.5$\times$ for 1024 neurons and a deceleration of 1.5$\times$ for 66K neurons.

\subsection{Variable Step Event Grouping}

On the analysis of the performance of the CVODE with grouping of events within half fixed timestep (2c$\hspace{0.3mm}{\color{gray}\bullet}$), when applied to the largest dataset tested, the previous acceleration was reduced to 47$\times$ for quiet, 4.4$\times$ for slow, and 1.2$\times$ for moderate dynamics, with an inferior performance on the remaining regimes. A further reduction of  speed-up to 33$\times$ for quiet and 1.9$\times$ for slow dynamics was noticeable on the CVODE implementation without events grouping (2c$\hspace{0.3mm}{\color{lightgray}\bullet}$), with lower performance for the remaining spike regimes. Although being more precise and solving correlated states implicitly, this method runs slower than the reference implicit fixed step method 2a$\hspace{0.3mm}{\color{blue}\bullet}$ in the use cases characterized by a high number of neurons and/or strong network activity. This goes in line with the conclusions in \sref{sec-in-vivo-exp}, confirming that performance is activity dependent, and the achievable speed-up depends on the network connectivity. %{\color{red}However, the situations of such performance decay are of rare occurrence or of short duration, therefore not being a blocker of performance in the overall simulation, and shown later.}

The speed-up introduced across FAP CVODE variants (2c$\hspace{0.3mm}{\color{black}\bullet}{\color{gray}\bullet}{\color{lightgray}\bullet}$) increases with the amount of discontinuities in the system --- correlated to high network activity or size --- as the efficiency of the event grouping method is related to the amount of events in the same grouping interval that are delivered at once. No difference in runtime was measured for the smallest dataset tested due to the high sparsity of synaptic events in small networks.

\begin{figure}[t]
\centering
\includegraphics[width=0.47\textwidth]{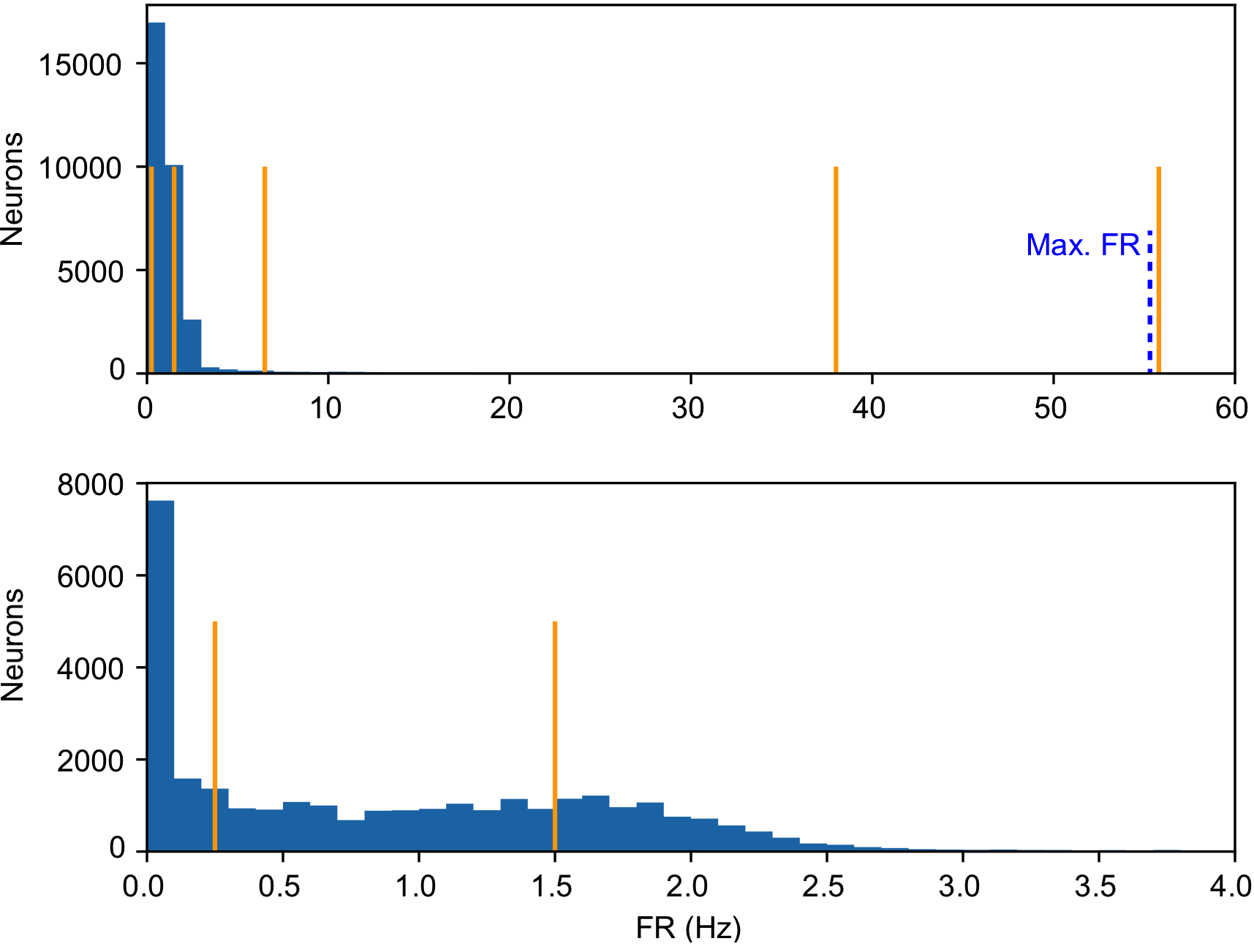} 
\caption{Distribution of neuron counts by spiking rates collected from the central minicolumn (31346 neurons) of the network utilized in the simulation of the laboratory experiment described in \sref{sec-in-vivo-exp}, in the interval 2-4 secs. Orange bars represent the frequencies  0.25 Hz, 1.5 Hz, 6.5 Hz, 38 Hz and 55 Hz that describe the boundaries of the 5 spiking regimes analysed. Top: a bin accounts for 1Hz. \textit{Max. FR} is the maximum firing rate measured; Bottom: zoomed dataset with a cut off at 0-40 Hz. A bin accounts for 0.1 Hz.}
\label{fig_spike_histogram}
\end{figure}

\subsection{Fully-Asynchronous vs Bulk-Synchonous Execution Models}

Our next analysis focuses on the performance difference between the BSP and FAP execution models. Results show that the runtimes of both implementations approximate with the increase of the input. This is visible by comparing the fixed step trajectories 1b\scalebox{0.8}{$\color{brown}\blacklozenge$} and 1c\scalebox{0.8}{$\color{cyan}\blacklozenge$}, and the variable step trajectories 2b$\hspace{0.3mm}{\color{red}\bullet}$ and 2c$\hspace{0.3mm}{\color{lightgray}\bullet}$. For small network sizes, the difference in runtime is noticeably few orders of magnitude higher than for larger network sizes. On large models, the runtimes are similar. This property was demonstrated in our previous work. In brief, an increase of network leads to a higher number of network connectivity, therefore reducing the maximum allowed stepping interval per neuron, and approximating it to the minimum communication delay utilised in the BSP methods. On fixed step methods, it is noticeable a similar runtime on large (66K) networks of neurons, as timesteps are computationally homogeneous. On variable step methods, similar runtimes are only noticeable when significant network activity is present (moderate, fast and burst dynamics), as little network activity leads to few discontinuities, allowing stepping intervals to be modelled in a reduced number of variable steps.

\subsection{Runtime Dependency on Input Size and Spike Activity}

\newcommand{\tikzsymbol}[2][circle]{\tikz[baseline=-0.5ex]\node[inner sep=2pt,shape=#1,draw,#2]{};}%

It is known that, on simulations of small networks, variable-timestep methods yield a significant acceleration in time to solution compared to fixed timestep methods \cite{lytton2005independent}. The rationale is that the small number of neurons in the network leads to a reduced number of synapses per neuron, thus less discontinuities.  However, the connectivity in larger networks reaches up to 10 thousand synapses per neuron, with the number of discontinuities being related also to the overall network activity. The question lies now on which conditions are required for similar computation complexity in both interpolators. To that extent, we  measured the regions of similar runtime growth for the reference fixed step (2b$\hspace{0.3mm}{\color{blue}\bullet}$) and our variable step methods (2c$\hspace{0.3mm}{\color{black}\bullet}$). The region is labelled as $\tikzsymbol[rectangle,orange]{minimum width=8pt,fill=orange!30}$ in \fref{fig_benchmark_xe6}.

As expected, fixed step methods yield a quasi-linear runtime growth with the increase of the input size, and are independent of the spiking regime, due to the almost ideal scaling of the algorithm in the BSP model.  On the other hand, the runtime of variable timestep methods --- dependent on the number of discontinuities ---  demonstrates a rapidly increasing growth with the input size outside the region of similar growth, and almost linearly inside. Moreover, as it depends on the network activity, the lower limit of the region increases with the spiking rate, and is delimited at 16.4K, 32.8K and 32.8K neurons or more for the quiet, slow and moderate dynamics, while not visible in the fast and burst dynamics. In the three spiking regimes where such region exists, the similar growth in both approaches provides a confidence of the scaling capabilities of our methods in larger network models. This is of high importance as it provides an estimation of runtime upper bound in simulations combining neurons with heterogeneous spiking rates, as discussed next.

\subsection{Overall Runtime Speed-up Estimation}

To conclude our analysis, we computed an estimation of the performance acceleration on a simulation combining several spiking regimes. For that purpose, we measured the the distribution of neuron spike rates and neurons per spiking regime, following the laboratory experiment simulation described in \sref{sec-in-vivo-exp}. Estimations were collected from the central minicolumn (31346 neurons) of the 219K neurons network, to avoid boundary-effects that would improve results due to neurons placed on the edges, with reduced connectivity.  The counts relate to the 2-4s simulation time interval, to exclude the initial artificial synaptic burst due to the current injection.  The results are presented in \fref{fig_spike_histogram}. The measured percentage of neurons on each regime is 31.43\%, 38.44\%, 27.02\%, 3.10\% and 0.01\%, relating to  68.9K, 84.3K, 59.2K, 6.8K and 22 neurons. Following the runtimes described in \sref{results_fixed_var}, the speed-up range for the interval of 1024-66K neurons when comparing our methods with the state-of-the-art solver for complex models (2a$\hspace{0.3mm}{\color{blue}\bullet}$) are estimated as:  224.5-11.9x for the variable step method with precise event delivery (2c$\hspace{0.3mm}{\color{lightgray}\bullet}$); 225.1-17.1x for the similar implementation with delivery of events within the next half timestep (2c$\hspace{0.3mm}{\color{gray}\bullet}$); and 228.5-24.6x for the use case with full-timestep event group delivery (2c$\hspace{0.3mm}{\color{black}\bullet}$).
%Calculated as, example:
%0.3143*(3754/6.96) + 0.3844*(3754/27.4) + 0.2702(3754/483) + 0.0310*(3754/1757) + 0.0001*(3754/2524) = 224.355357428
%As mentioned previously, the fast and burst dynamics weight lightly in the overall runtime, due to their little occurrence. On the other hand, the quiet, slow and moderate dynamics drive the performance increase of our methods.
 As a side note, the percentage of neurons in the quiet dynamics regime does not agree with the 90\% described in the literature. We believe this is due to the reduced size of the network, and that simulations of larger networks and complete brain models include a larger portion of neurons in this regime.  Therefore, we assume these results to be a lower bound estimate of possible acceleration.

Moreover, since the quiet, slow and moderate dynamics regimes weight over 95\% in the runtime calculation, and as for datasets above 32.8K the reference vs benchmark runtimes have a similar runtime growth in those regimes, we believe the overall runtime for larger circuits do not yield a significant reduction in the speed-up values presented, and that the scaling properties are almost fully-preserved on larger networks.

\section*{Discussion}

This paper presented a strategy for the fully-asynchronous distributed simulation of detailed neuron models with variable-order variable-timestep interpolation. We detailed state-of-the-art approaches based on the Bulk Synchronous Parallel execution model (BSP), their limitations on the numerical resolution of complex neuron models, and computation load imbalance at synchronization barriers in variable-step simulations. We discussed the problem of synaptic exchange and synchronization on networks of neurons on speculative variable-step executions, and the inherent issue of a cascade chain of states reset and reversal of false synaptic events. To overcome this issue, we proposed an alternative non-speculative scheduled execution model on a  distributed network of compute nodes. The approach follows a novel Fully-Asynchronous Parallel (FAP) execution model (yielding asynchronous computation, communication and synchronisation), that relies on the individual stepping of neurons based on the time instant of their synaptic connectivities, avoiding backstepping and allowing for stepping intervals beyond the BSP-based synchronization interval. Step lengths are maximised by a scheduler that tracks neurons' time advancement and dynamically allocates compute resources to the earliest neurons in time. 

We performed an analysis of numerical accuracy and demonstrated better precision and less interpolation steps of the methods presented compared to the reference fixed-step implicit solver in the NEURON scientific application. An analysis of variable-step performance based on step count and time to solution was performed on three experimental set-ups. (1) A continuous current injection causing fast gradient changes in state on a single neuron demonstrated a reduced step count and runtime, and low dependency of performance on the stiffness of solution. (2) An injection of a sequence of current pulses simulating network activity demonstrated a high dependency of step count and runtime on incoming synaptic activity, which cause a solution discontinuity and a reset of solution state. We demonstrated a reduced step count for an incoming spike current rates between 1000 and 1600 Hz, depending on the current value. (3) A digital reconstruction of a laboratory experiment on a network of 219 thousand neurons measured the relevance of our result to a real use case. Results displayed a highly heterogeneous distribution of discontinuity rates across neurons, demonstrating the suitability of our methods to the problem domain.

A proof of concept simulator was implemented on top of the core kernel of the NEURON scientific application. Distributed asynchrony, global memory addressing space, remote procedure calls, threading, synchronization and communication methods were replaced and implemented by the HPX runtime system \cite{sterling14:_easc}. %Variable step resolution is provided by CVODE \cite{cohen1996cvode}, an implementation of the Backward Differentiation Formula for the variable-step multistep implicit method solving the Initial Value Problem. 
We analysed and benchmarked five spiking regimes that describe distinct patterns of brain activity in mammals, and six state-of-the-art solvers for simple and complex neuron models. Benchmark results on a network of 1024-65536 neurons demonstrate an overall reduction in runtime in the order of 224.5-11.9x for the most accurate method with precise delivery of events in a continuous timeline, and 225.1-17.1x and 228.5-24.6x for two optimized variants with half-step and full-step grouped delivery of events. We demonstrated that performance is activity-dependent and that almost ideal scaling is possible, as over 95\% of neurons in a biologically-inspired neuron network --- from a simulated laboratory experiment --- fall in spiking regimes with guaranteed preservation of the speed-up achieved.

The fully-asynchronous variable-step methods presented open the prospectus for the redesign of simulations across a wide range of scientific domains, so far limited to synchronous fixed-step interpolation on the BSP execution model. Furthermore, the scaling, accuracy and performance demonstrated can be further improved with enhanced numerical resolution and asynchronous computation. 
Thus, as future work, we intend to study the simulation of further laboratory experiments, a lower-order BDF for reduced memory usage, fine-tuned tolerance values, a mechanism for the \textit{fallback} to standard Backward Euler for neurons in fast and burst spiking regimes, and a \textit{carefully-speculative} execution model performing speculative stepping while avoiding the initiation of cascades of solution resets throughout neurons.  Moreover, we plan to investigate dynamic load balancing of neurons across compute nodes based on recent momentary activity. Combined with the FAP execution model presented, it opens new prospectus in solving the long-standing problem of computational imbalance at synchronization barriers that characterizes BSP-based variable-step simulations.

\section*{Acknowledgements}

The work was supported by funding from the ETH Domain for the Blue Brain Project (BBP). The super-computing infrastructures were provided by Indiana University. A portion of Michael Hines efforts was supported by NINDS grant R01NS11613. The authors would like to thank Francesco Cremonesi for scientific discussions and Max Nolte for the provision of the spiking rate information.

\bibliography{neurox}

\end{document}